\documentclass{article}
\usepackage[utf8]{inputenc}
\usepackage{rotating}
\usepackage{graphicx}
\usepackage{amsmath}
\usepackage{caption,subcaption}
\usepackage[export]{adjustbox}
\usepackage{soul}
\usepackage{comment}
\usepackage{multirow}
\usepackage{xcolor}
\usepackage{authblk}
\usepackage{booktabs}
\usepackage{bbold}
\usepackage{relsize}
\usepackage{authblk}
\usepackage[symbol]{footmisc}
\renewcommand{\thefootnote}{\fnsymbol{footnote}}
\usepackage{hyperref}
\usepackage{array}
\usepackage{algorithm}
\usepackage{algorithmic}
\usepackage{longtable}
\usepackage[left=2cm, right=2cm, top=2cm]{geometry}
\usepackage{framed}
\usepackage[
backend=biber,
style=numeric,
sorting=none
]{biblatex}
\usepackage[normalem]{ulem}
\usepackage{array}
\usepackage[title]{appendix}
\newcolumntype{P}[1]{>{\centering\arraybackslash}p{#1}}

\usepackage{tikz}
\usetikzlibrary{shapes.geometric, arrows}

\tikzstyle{startstop} = [rectangle, rounded corners, minimum width=3cm, minimum height=1cm,text centered, draw=black, fill=red!30]
\tikzstyle{process} = [rectangle, minimum width=3cm, minimum height=1cm, text centered, draw=black, fill=blue!10]
\tikzstyle{decision} = [diamond, minimum width=3cm, minimum height=1cm, text centered, draw=black, fill=green!20]
\tikzstyle{arrow} = [thick,->,>=stealth]

\newcommand{\revisions}[1]{#1}

\title{Link Statistics of Dislocation Network during Strain Hardening}

\author[1,*]{Sh.\ Akhondzadeh}
\author[1,*]{Hanfeng Zhai}
\author[1]{Wurong Jian}
\affil[1]{Department of Mechanical Engineering, Stanford University, Stanford, CA, 94305, USA}
\author[2]{Ryan B. Sills}
\affil[2]{Department of Materials Science and Engineering, Rutgers University, Piscataway, NJ, 08854, USA}
\author[3]{Nicolas Bertin}
\affil[3]{Lawrence Livermore National Laboratory, Livermore, CA, 94550, USA}
\author[1]{Wei Cai\footnote{E-mail: \texttt{caiwei@stanford.edu}}}
\affil[*]{\rm These authors contributed equally to this work.}

\begin{document}
\maketitle
\renewcommand{\thefootnote}{\arabic{footnote}}
\setcounter{footnote}{0}

\begin{abstract} 
Dislocations are line defects in crystals that multiply and self-organize into a complex network during strain hardening. The length of dislocation links, connecting neighboring nodes within this network, contains crucial information about the evolving dislocation microstructure.
%
By analyzing data from Discrete Dislocation Dynamics (DDD) simulations in face-centered cubic (fcc) Cu, we characterize the statistical distribution of link lengths of dislocation networks during strain hardening on individual slip systems. 
Our analysis reveals that link lengths on active slip systems follow a double-exponential distribution, while those on inactive slip systems conform to a single-exponential distribution. 
The distinctive long tail observed in the double-exponential distribution is attributed to the stress-induced bowing out of long links on active slip systems, a feature that disappears upon removal of the applied stress. 
We further demonstrate that both observed link length distributions can be explained by extending a one-dimensional Poisson process to include different growth functions.
%
%
Specifically, the double-exponential distribution emerges when the growth rate for links exceeding a critical length becomes super-linear, which aligns with the physical phenomenon of long links bowing out under stress. 
This work advances our understanding of dislocation microstructure evolution during strain hardening and elucidates the underlying physical mechanisms governing its formation.
%
%
%
%
%
%
\end{abstract}


\newpage

\section{Introduction}
\label{sec:intro}

Dislocations, the line defects that define the boundary between slipped and unslipped regions in crystalline materials, are the dominant carriers for plastic deformation under most conditions. 
Although the role of dislocations in crystal plasticity has been recognized for over 80 years~\cite{taylor1934mechanism, orowan1934plasticity, polanyi1934lattice}, a quantitative link between the dynamics of individual dislocations and the macroscopic stress-strain behavior of crystalline materials has been lacking.
%
This gap is in part due to the complex microstructure (i.e.~network) that the dislocations self-organize into during plastic deformation~\cite{MT2020}.
Quantifying the dislocation microstructure is an essential step towards establishing a physics-based constitutive model for crystal plasticity~\cite{zaiser2019stochastic,sills2018dislocation}.
%
%


The dislocation microstructure is a network of nodes interconnected by dislocation links. 
Each link is a dislocation line with the same Burgers vector and slip plane, connecting at both ends to segments belonging to different slip systems.
The length distribution of the dislocation links is a fundamental statistical property of the dislocation microstructure~\cite{Arechabaleta2018_MSEA}.
Given its importance, the dislocation link length distribution has been investigated both experimentally (through electron microscopy) and theoretically~\cite{ajaja1986dislocation, ardell1984dislocation, lagneborg1973model, modeer1974dislocation, shi1993dislocation, Lin1989_acta}.
%
However, most of these studies have focused on metals deformed under high-temperature creep conditions, instead of during strain-hardening under ambient conditions, which is our focus here.
Recently, Sills et al.~\cite{sills2018dislocation} investigated the dislocation link length distribution during strain-hardening in Discrete Dislocation Dynamics (DDD) simulations of single-crystal copper deformed along the $[0\,0\,1]$ orientation.
They reported that the link lengths follow an exponential distribution, which can be completely specified by two parameters (the total number of links $N$ and the average link length $\bar{l}$).
%
%
Sills et al.'s analysis~\cite{sills2018dislocation} considered the links on all the slip systems together.
This is a reasonable approximation for the high-symmetry, $[0\,0\,1]$ loading direction, where 8 out of the 12 slip systems are equally active and the remaining 4 slip systems are inactive.
However, for general loading directions, significant variations in slip system activity are expected, making it important to consider link length distributions on individual slip systems.

%

In this work, we analyze the link length distribution on individual slip systems, using data from our previous DDD strain-hardening simulations for uniaxial loading along 118 crystallographic directions~\cite{JMPS1}.
%
%
%
%
%
%
%
Compared to Sills et al.~\cite{sills2018dislocation}, we examine more configurations to improve the statistics and extract link distributions over a wider range of link lengths.
%
%
We show that during plastic deformation, the link length distributions on active slip systems can often be described by a sum of two exponential functions,
\begin{equation}\label{eq:2Expo_i}
    n_i(l) = \frac{N_i^{(1)}}{\bar{l}_i^{(1)} } \, e^{-l / \bar{l}_i^{(1)} }
           + \frac{N_i^{(2)}}{\bar{l}_i^{(2)} } \, e^{-l / \bar{l}_i^{(2)} }
           \, ,
\end{equation}
where $i=1, \cdots, 12$ is the slip-system index, and $N_i^{(1)}$, $N_i^{(2)}$, $\bar{l}_i^{(1)}$, $\bar{l}_i^{(2)}$ are the four parameters characterizing this distribution.
The first exponential (dominating for link lengths up to $5\,\bar{l}_i$, where $\bar{l}_i$ is the averaged link length on slip system $i$) captures most links and aligns with previous findings~\cite{sills2018dislocation}, while the second exponential corresponds to much longer, links of high velocity (dominating for link lengths from $5\,\bar{l}_i$ to $25\,\bar{l}_i$) \cite{akhondzadeh2021statistical}.
%
On the other hand, the link length distribution on inactive slip systems can still be well described by a single exponential function.
%
%
If the deformation process is interrupted, and the sample is relaxed to the zero-stress state, then we find that a single exponential function can well describe the link length distribution on all slip systems, consistent with~\cite{sills2018dislocation}.
%
We also report the general trend of dependence of the characteristics of the link length distributions on the loading orientations in the stereographic triangle.
For loading orientations near the corners of the stereographic triangle, a strong long-tail in the double-exponential distribution is observed on slip systems $i$ with a moderately high Schmid factor $S_i$ (e.g. $S_i > 0.2$).
%
%
However, near the central region of the stereographic triangle, the long-tail in the double-exponential distribution is quite weak, even for slip systems with very high Schmid factors (e.g. $S_i > 0.3)$.

To explain the observed single and double exponential distributions for link lengths, we developed a model that generalizes the one-dimensional Poisson process, in which links are randomly split into shorter links
with a probability rate, by also allowing the links to grow in length.
%
%
When the growth rate $\dot{l}$ is zero or linear with link length $l$, this model reproduces the single-exponential link length distribution, consistent with~\cite{sills2018dislocation}.
If the growth rate $\dot{l}$ exceeds the linear function for longer links, then the model reproduces the double-exponential distribution.
These findings suggest that the double-exponential distribution observed in the DDD simulations is due to the fact that during plastic deformation, longer links bow out and expand much faster than shorter links~\cite{strm1979, Lagneborg1972, KuhlmannWilsdorf1989_MSEA}.
We also provide numerical evidence from DDD simulations that on active slip systems, long links indeed move at much higher velocities than short links on the same slip system.

The remainder of this paper is organized as follows. 
In Section~\ref{sec:methods}, we explain how the data from DDD simulations is extracted and analyzed.
In Section~\ref{sec:results}, we demonstrate that the link length distributions for individual slip systems often follow a single‐ or double‐exponential form. 
%
We examine the dependence of the link length distribution on loading orientation by showing four typical examples.  We also show the distribution of the fitted statistical parameters for all the loading orientations considered.
%
%
In Section~\ref{sec:discussion}, we introduce a generalized Poisson process model that sheds light on the physical origin of single- and double-exponential link length distributions.
We also provide evidence from DDD simulations that longer links move faster than shorter links.
%
%
%
%
%
In Section~\ref {sec:conclusion}, we provide some concluding remarks.

%




\section{Methods}
\label{sec:methods}
%

\begin{figure}[htbp]
    \centering
    \includegraphics[width=0.9\linewidth]{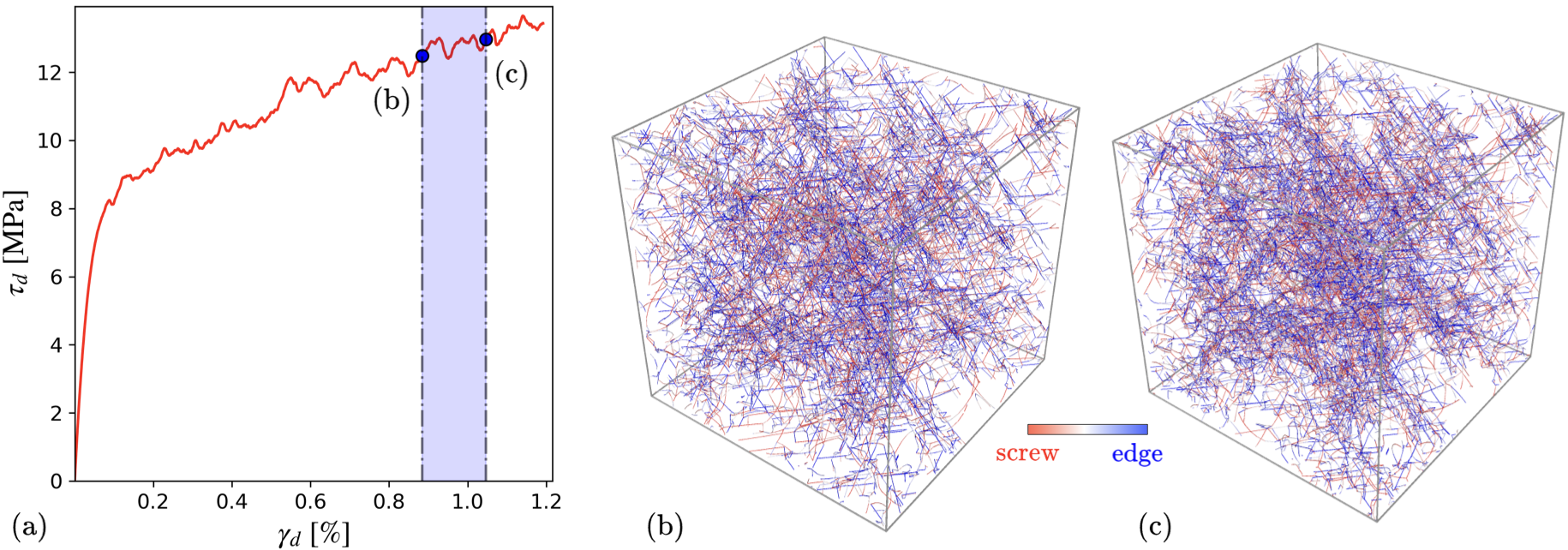}
    \caption{Schematic illustration for accumulating the link length statistics based on the loading direction $[0.03,\ 0.05,\ 0.99]$. (a) The resolved shear stress-strain curve. The strain region $\gamma_d\in[0.9,\ 1.05]\%$, over which the link length statistics are averaged, are shown in the shaded area. The dislocation structures corresponding to the start and end of the averaging window are shown in (b) and (c), respectively. The dislocation lines are colored based on their character angle (red for screw and blue for edge orientations).}
    \label{loading schematic}
\end{figure}

We used data from previous discrete dislocation dynamics (DDD) simulations~\cite{JMPS1} performed with the ParaDiS code~\cite{Arsenlis2007_msmse}.
These simulations were conducted for copper (Cu), using the material properties listed in Table~\ref{tbl:simParam}.
The simulations were initialized with randomly distributed dislocation lines in a \((15\,\mu\mathrm{m})^3\) cell and deformed to various strain levels under different loading orientations in the stereographic triangle.
Figure~\ref{loading schematic}(a) shows the shear stress–strain curve for a loading direction close to $[0\, 0\, 1]$.
Figure~\ref{loading schematic}(b) and (c) show dislocation snapshots at two strain levels well inside the strain-hardening regime.

To analyze dislocation link lengths, we begin with a given dislocation configuration (e.g., Figure~\ref{loading schematic}(b)) and extract all dislocation links, defined as a sequence of dislocation segments sharing the same Burgers vector and glide plane.  
Each configuration contains approximately \(2\times10^4\) links.
We then construct a histogram of link lengths $l$, normalized by the average link length $\overline{l}_i$ for that configuration.
The histogram is constructed over 125 regularly spaced bins on the domain of $l/\overline{l}_i$ from 0 to 25.
To enhance statistical significance, we average these normalized histograms across all configurations --- each corresponding to a time step of the DDD simulation --- within a defined strain window (e.g., [0.9, 1.05]\%).  
%
%
This approach allows us to capture link length density data $n_i(l)$ spanning six orders of magnitude, significantly exceeding the range attainable from a single configuration~\cite{sills2018dislocation}.  
%
%
%
The selection of the strain window involves balancing two key considerations:  sufficient width for adequate statistical sampling and the requirement that the dislocation network structure remains relatively consistent throughout the window, as illustrated, for example, by Figures~\ref{loading schematic}(b) and (c).
%
%

When the link length distribution on a given slip system $i$ clearly deviates from a single exponential form (i.e. a straight line when plotted using the semi-log-$y$ scale), we fit it to the double-exponential form of Eq.~\eqref{eq:2Expo_i}.
Because the total number of links $N_i$ and the averaged link length $\overline{l}_i$ can be easily calculated from the DDD data, they are not considered fitted parameters.
Instead, they provide constraints on the four parameters that characterize the length distribution, $N_i^{(1)}$, $N_i^{(2)}$, $\bar{l}_i^{(1)}$, $\bar{l}_i^{(2)}$, in the form of: $N_i = N_i^{(1)} + N_i^{(2)}$ and $N_i\, \overline{l}_i = N_i^{(1)}\, \overline{l}_i^{(1)} + N_i^{(2)}\, \overline{l}_i^{(2)}$.
We find it convenient to introduce two dimensionless coefficients: 
$\lambda_i^{(1)}\equiv \bar{l}_i^{(1)} \,/\, \bar{l}_i$ and $\lambda_i^{(2)}\equiv \bar{l}_i^{(2)} \,/\, \bar{l}_i$, in terms of which 
we can
express $N_i^{(1)}$, $N_i^{(2)}$, $\bar{l}_i^{(1)}$, $\bar{l}_i^{(2)}$ with the two constraints above automatically satisfied.
%
We obtain $\lambda_i^{(1)}$ and $\lambda_i^{(2)}$ by fitting to the DDD data using a nonlinear least squares method.
%
We employ the Trust Region Reflective algorithm~\cite{virtanen2020scipy} to minimize the discrepancy between $\ln\left(n_i(l_{\rm bin})\right)$ with $n_i(l)$ extracted from DDD data and those with $n_i(l)$ expressed by Eq.~\eqref{eq:2Expo_i}. 
%
%
Here, $l_\text{bin}$ is the link length corresponding to the center of each bin of the histogram.
%
%
%
%
%

\section{Results}
\label{sec:results}

\begin{figure}[htbp]
    \centering 
    \includegraphics[width=0.35\linewidth]{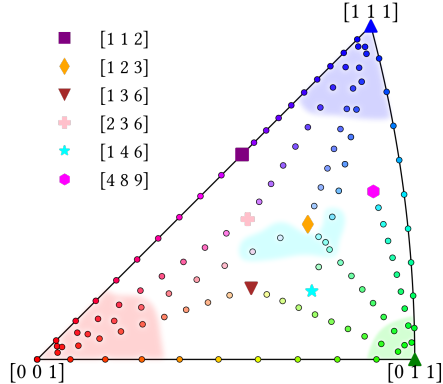}
    \caption{The stereographic triangle with symbols corresponding to loading orientations. Circles correspond to the loading orientations of the 118 DDD simulations used in this analysis.
    Other symbols correspond to reference orientations used to define the loading orientations for the DDD simulations.
    The four shaded regions correspond to areas where typical behaviors are reported in Sections~3.1 to 3.4.
    %
    }\label{fig:overall_stereo_behavior}
\end{figure}


DDD simulations were performed with 118 loading orientations as shown in Figure~\ref{fig:overall_stereo_behavior}.  They correspond to circles in the stereographic triangle, and are the linear interpolations between special orientations marked by other symbols.
%
%
The symbols are colored using
the Electron Backscatter Diffraction (EBSD) coloring scheme.
In Figure~\ref{fig:overall_stereo_behavior}, we also marked four shaded regions that exhibit typical behaviors of the link length distribution, which we will report in the following subsections.
%
%
%


\subsection{\texorpdfstring{Loading axis near $[0\,0\,1]$}{Loading axis near [0 0 1]}}
\label{sec:near_001}

\begin{figure}[htbp]
\includegraphics[scale=1,center]{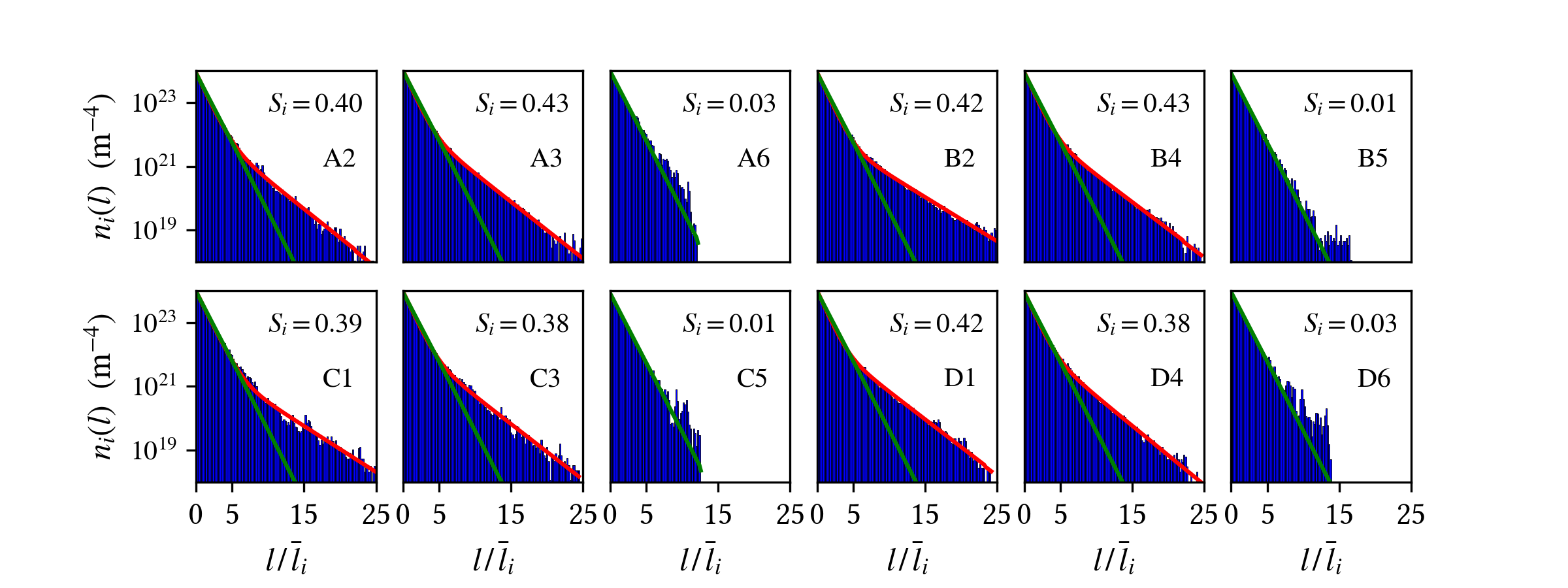}
\caption[justification=centering]{Link length distribution of dislocations on individual slip systems corresponding to a DDD simulation along $[0.03,\,0.05,\,0.99]$ loading orientation, averaged over the strain window of $\gamma_d \in 0.9-1.05\%$. For active slip systems, the red line represents the best fit to the DDD data using two exponentials as shown in Eq.~(\ref{eq:2Expo_i}), the green line indicates the fit corresponding to the first term in the double-exponential distribution; For inactivate slip systems, the green line is the single exponential function shown in Eq.~(\ref{eq:1Expo_i}).}
\label{fig:LL_system}
\end{figure}

Figure~\ref{fig:LL_system} shows the link length distribution of $n_i(l)$ corresponding to loading orientation $[0.03, 0.05, 0.99]$
on the 12 slip systems, labeled by the Schmid–Boas (SB) notation (see Table~\ref{tab:2exp_ddd_fit}). 
The observed behavior is typical of the loading orientations shown in the shaded area near the $[0\,0\,1]$ corner in Figure~\ref{fig:overall_stereo_behavior}.
For slip systems A6, B5, C5, D6, $n_i(l)$ appears to follow a single-exponential distribution, i.e., appearing as a straight line on a semilogarithmic plot --- consistent with earlier observations~\cite{sills2018dislocation}.
%
%
%
%
%
The link length distribution on these four slip systems can be described by a single-exponential function (with no fitting parameters),
\begin{equation}\label{eq:1Expo_i}
    n_i^{\rm S}(l) = \frac{N_i}{\bar{l}_i } \, e^{-l / \bar{l}_i } \, ,
\end{equation}
%
%
where $N_i$ is the total number of links and $\bar{l}_i$ is the average link length. Both $N_i$ and $\bar{l}_i$ can be easily calculated from the collection of links without parameter fitting.
%

In contrast, the remaining slip systems exhibit link length distributions that require a double-exponential function, as defined in Eq.~\eqref{eq:2Expo_i}, characterized by four parameters $N_i^{(1)}$, $N_i^{(2)}$, $\bar{l}_i^{(1)}$, $\bar{l}_i^{(2)}$.
%
As described in Section~\ref{sec:methods}, the parameters
%
are constrained by $N_i^{(1)} + N_i^{(2)} = N_i$, and $N_i^{(1)}\,\bar{l}_i^{(1)} + N_i^{(2)}\,\bar{l}_i^{(1)} = N_i\,\bar{l}_i$, 
and are expressed in terms of two dimensionless parameters $\lambda_i^{(1)}$ and $\lambda_i^{(2)}$.
%
%

\begin{table}[htbp]
\centering
\begin{tabular}{c c c c c c c c c c c c}
\toprule
Index $i$ & $\mathbf{n}_i$ & $\mathbf{b}_i$ & SB index & $S_i$ & $N_i$ & $N_i^{(1)}$ & $N_i^{(2)}$ & $\bar{l}_i$ & $\lambda_i^{(1)}$ & $\lambda_i^{(2)}$ & $\rho_i$ \\ 
\midrule
1  & $(\bar{1}11)$ & $\tfrac{1}{2}[0\bar{1}1]$ & A2 & $0.40$ & $4.15$ & $3.89$ & $0.26$ & $4.53$ & $0.91$ & $2.42$ &  $1.88$ \\ 
2  & $(\bar{1}11)$ & $\tfrac{1}{2}[101]$       & A3 & $0.43$ & $5.64$ & $5.15$ & $0.48$ & $5.32$ & $0.86$ & $2.44$ & $3.00$ \\
3  & $(\bar{1}11)$ & $\tfrac{1}{2}[110]$       & A6 & $0.03$ & $4.17$ & ---    & ---    & $3.94$ & ---    & ---    & $1.64$ \\
4  & $(111)$       & $\tfrac{1}{2}[0\bar{1}1]$ & B2 & $0.42$ & $5.06$ & $4.83$ & $0.23$ & $5.43$ & $0.90$ & $3.13$ & $2.75$ \\
5  & $(111)$       & $\tfrac{1}{2}[\bar{1}01]$ & B4 & $0.43$ & $4.95$ & $4.57$ & $0.37$ & $5.16$ & $0.87$ & $2.55$ & $2.55$ \\
6  & $(111)$       & $\tfrac{1}{2}[\bar{1}10]$ & B5 & $0.01$ & $3.83$ & ---    & ---    & $4.26$ & ---    & ---    & $1.63$ \\
7  & $(\bar{1}11)$ & $\tfrac{1}{2}[011]$       & C1 & $0.39$ & $4.86$ & $4.74$ & $0.12$ & $4.63$ & $0.95$ & $3.01$ & $2.25$ \\
8  & $(\bar{1}11)$ & $\tfrac{1}{2}[101]$       & C3 & $0.38$ & $4.25$ & $3.94$ & $0.31$ & $4.29$ & $0.88$ & $2.48$ & $1.83$ \\
9  & $(\bar{1}11)$ & $\tfrac{1}{2}[110]$       & C5 & $0.01$ & $3.72$ & ---    & ---    & $4.01$ & ---    & ---    & $1.49$ \\
10 & $(\bar{1}11)$ & $\tfrac{1}{2}[011]$       & D1 & $0.42$ & $5.38$ & $4.87$ & $0.50$ & $5.17$ & $0.85$ & $2.50$ & $2.78$ \\
11 & $(\bar{1}11)$ & $\tfrac{1}{2}[\bar{1}01]$ & D4 & $0.38$ & $4.57$ & $4.16$ & $0.40$ & $4.68$ & $0.87$ & $2.35$ & $2.14$ \\
12 & $(\bar{1}11)$ & $\tfrac{1}{2}[110]$       & D6 & $0.03$ & $3.89$ & ---    & ---    & $3.88$ & ---    & ---    & $1.51$ \\ 
\bottomrule
\end{tabular}
\caption{Slip systems in FCC crystals (Schmid factors for loading $[0.02,\ 0.03,\ 0.99]$) with their plane normal $\mathbf{n}_i$ and Burgers vector $\mathbf{b}_i$, and their Schmid-Boas (SB) index.
All $N_i$, $N_i^{(1)}$, and $N_i^{(2)}$ entries are in units of $10^{17}\,\mathrm{m}^{-3}$, $\bar l_i$ in $10^{-7}\,\mathrm{m}$, and $\rho_i$ in $10^{11}\,\mathrm{m}^{-2}$.}
\label{tab:2exp_ddd_fit}
\end{table}

The double-exponential distribution suggests that the dislocation links on active slip systems can be considered as consisting of two populations: (1) and (2), each having its own number of links $N_i^{(1)}$ or $N_i^{(2)}$ and its average link length $\bar{l}_i^{(1)}$ or $\bar{l}_i^{(2)}$.
The fitted parameters $N_i^{(1)}$, $ N_i^{(2)}$, $\lambda_i^{(1)}$, $\lambda_i^{(2)}$, together with $N_i$ and $\bar{l}_i$ on all slip systems for the loading orientation $[0.03, 0.05, 0.99]$ are listed in Table~\ref{tab:2exp_ddd_fit}.
We note that $N_i^{(2)} \ll N_i^{(1)}$, and $\lambda_i^{(1)} \approx 1$, meaning that the population (1) contains most of the links, while the population (2) is the minority and contains the long tail of the distribution.
%
%
%
Consequently, the exponential distribution (Eq.~\eqref{eq:1Expo_i} or the first term in Eq.~\eqref{eq:2Expo_i}) accurately describes the link length distribution for link lengths between $0$ and $5\,\bar{l}_i$ for all slip systems, regardless of whether the overall distribution is a single or double-exponential.

Figure~\ref{fig:LL_system} shows that the link lengths on active slip systems satisfy a double exponential distribution and those on inactive slip systems satisfy a single exponential distribution.
In this loading orientation (and those in the shaded area near $[0\,0\,1]$ in Figure~\ref{fig:overall_stereo_behavior}), the active and inactive slip systems can be very easily distinguished by their Schmid factors $S_i$.
Here, the Schmid factors of active slip systems are all above 0.3 and those of inactive slip systems are all below 0.1.
%
%
%
%
%
%
We hypothesize that the long tail in the double-exponential link length distribution is caused by the bowing out of the links on the active slip system, driven by the resolved shear stress.
The absence of the long tail in the link length distribution on inactive slip systems is due to the inability of the links to bow out due to the insufficient resolved shear stress.

%
%
%

\begin{figure}[htbp]
\includegraphics[scale=1.0,center]{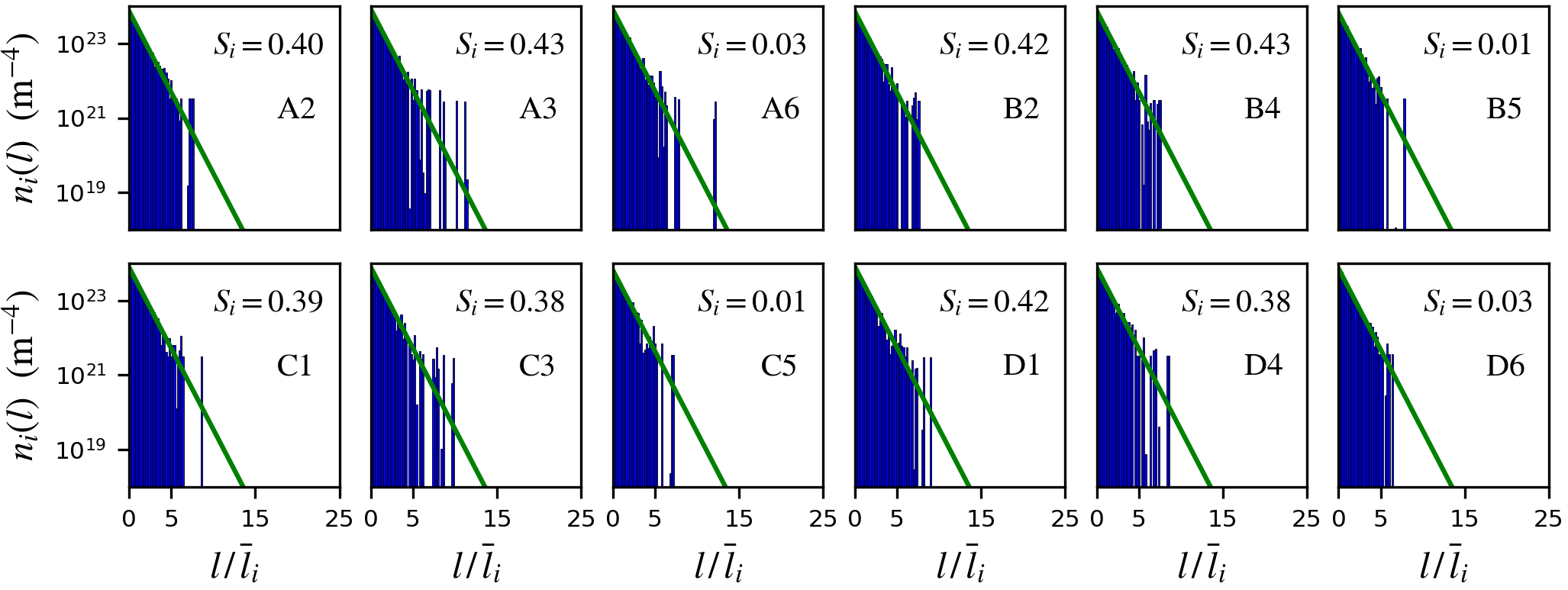}
\caption[justification=centering]{Link length distribution of dislocations on individual slip systems corresponding to a relaxed configuration taken at $\gamma_d=0.9\%$ under $[0.03,\,0.05,\,0.99]$ loading orientation. Single exponential is shown by green lines. The fitting coefficient of the single exponential is imposed to be 1.
}
\label{fig:LL_system_xed}
\end{figure}

To test this hypothesis, we relax the dislocation configuration at 0.9\% shear strain (shown in Figure~\ref{loading schematic}(b)) to zero stress.
%
%
%
Figure~\ref{fig:LL_system_xed} shows the link length distribution of the relaxed configuration, which can be well described by single-exponential distributions on all slip systems, consistent with Ref.~\cite{sills2018dislocation}.
%
This result confirms our hypothesis that the bowing out of dislocation links driven by the applied stress is the cause of the long tail in the double-exponential distribution on active slip systems. We will provide direct evidence on the link velocity in Section~\ref{sec:link_velocity}.
%
%
%
This finding also means that the long tail in the double-exponential can be observed only by \emph{in situ} experiments during the plastic deformation.  We expect the link length distribution to become single-exponential in \emph{ex situ} experiments when the applied stress has been removed.
%


\subsection{\texorpdfstring{Loading axis near $[0\,1\,1]$}{Loading axis near [0 1 1]}}
\label{sec:near_011}
\begin{figure}[htbp]
\includegraphics[scale=1,center]{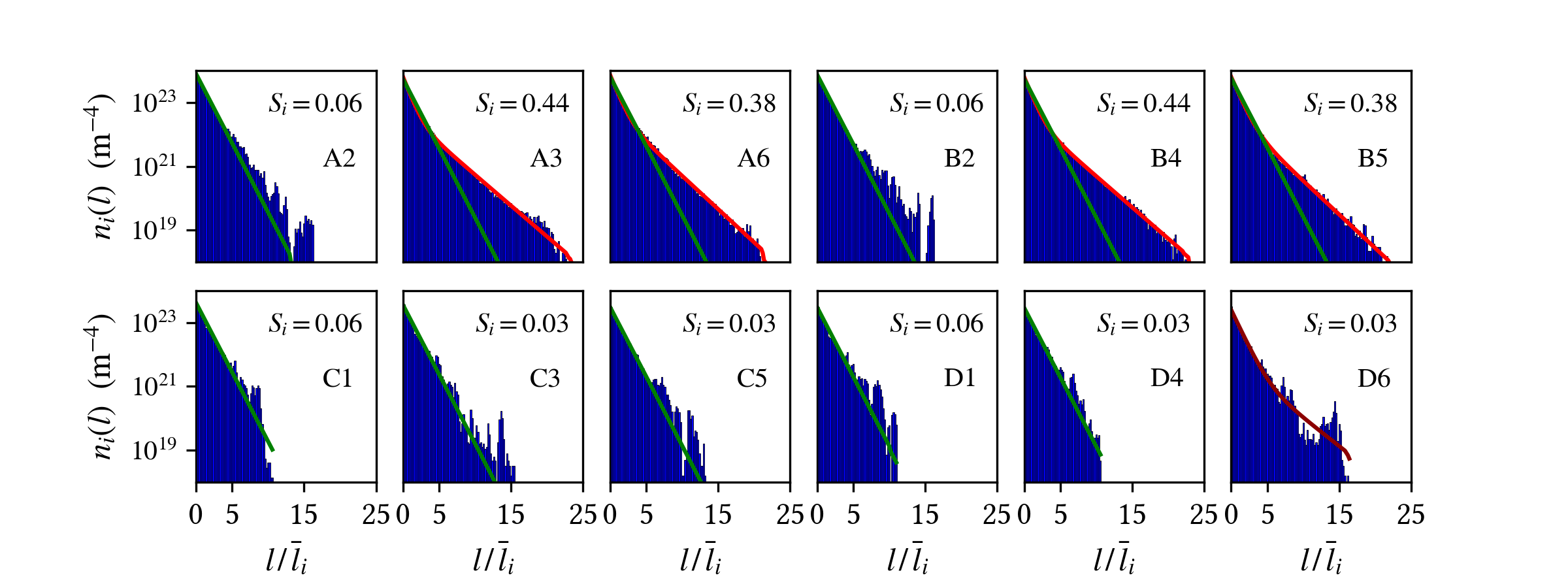}
\caption[justification=centering]{Link length distribution of dislocations on individual slip systems corresponding to a DDD simulation along $[0.00,\,0.65,\,0.76]$ loading orientation.}
\label{fig:LL_system_011}
\end{figure}

%
%
%
%

Figure~\ref{fig:LL_system_011} shows the link length distribution of $n_i(l)$ corresponding to loading orientation $[0.00, 0.65, 0.76]$
on the 12 slip systems. 
The observed behavior is typical of the loading orientations shown in the shaded area near the $[0\,1\,1]$ corner in Figure~\ref{fig:overall_stereo_behavior}.
Figure~\ref{fig:LL_system_011} shows that
%
slip systems A3, A6, B4, and B5 exhibit double-exponential link length distributions, while the remaining 8 slip systems exhibit (close-to) single-exponential distributions.
Here, the slip systems with double-exponential distributions can be identified as active slip systems (with Schmid factor $> 0.3$) and the slip systems with single-exponential distributions can be identified as inactive slip systems (with Schmid factor $< 0.1$).
This finding further supports our hypothesis that the long-tail in the double-exponential link length distribution is caused by bowing out of long links on active slip systems due to the resolved shear stress.

We note that slip system D6 (with Schmid factor 0.03) exhibits a visible deviation from single-exponential distribution, but it does not exhibit a clear double-exponential distribution either.
We classify this kind of slip systems as not clearly obeying either single- or double-exponential link length distributions (with fitted curve colored brown).

%
%
%
%
%

\subsection{\texorpdfstring{Loading axis near $[1\,1\,1]$}{Loading axis near [1 1 1]}}
\label{sec:near_111}

\begin{figure}[htbp]
\includegraphics[scale=1,center]{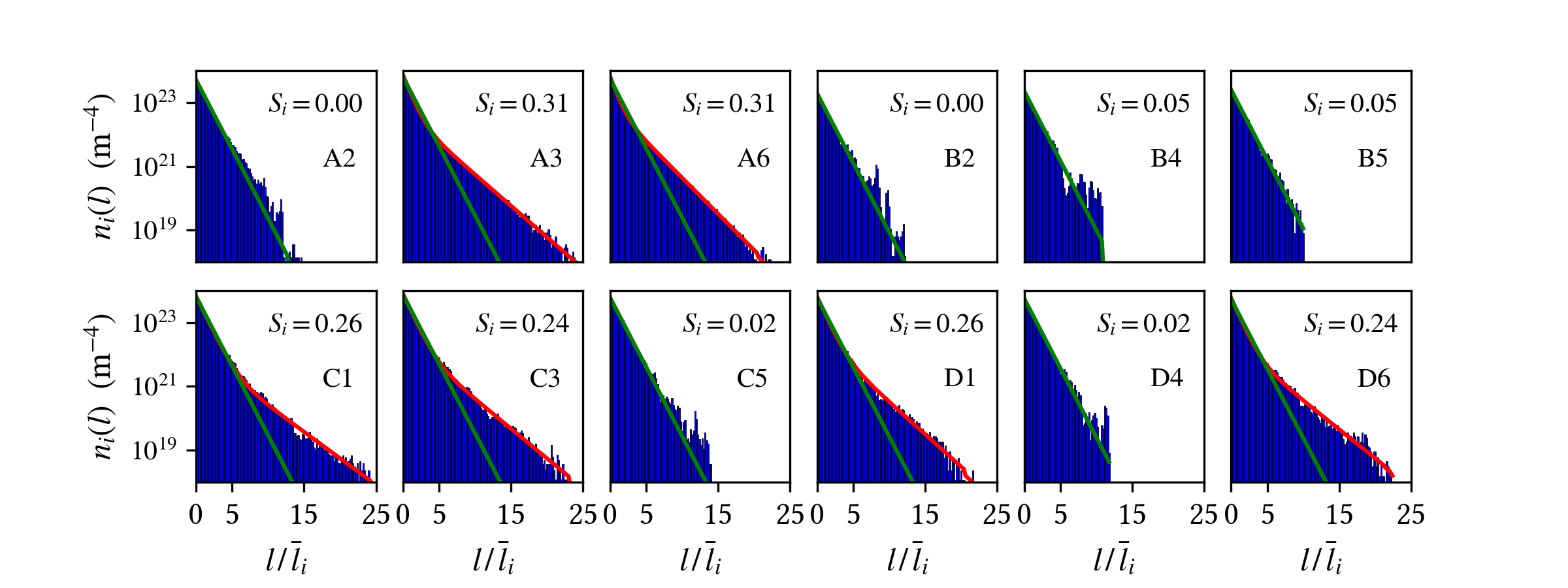}
\caption[justification=centering]{Link length distribution of dislocations on individual slip systems corresponding to a DDD simulation along $[0.53,\,0.60,\,0.60]$ loading orientation.}
\label{fig:LL_system_111}
\end{figure}

Figure~\ref{fig:LL_system_111} shows the link length distribution of $n_i(l)$ corresponding to loading orientation $[0.53, 0.60, 0.60]$
on the 12 slip systems. 
The observed behavior is typical of the loading orientations shown in the shaded area near the $[1\,1\,1]$ corner in Figure~\ref{fig:overall_stereo_behavior}.
%
Double-exponential link length distributions are observed on active slip systems, A3, A6, C1, C3, D1, D6 (with Schmid factors $> 0.2$), while single-exponential link length distributions are observed on the remaining, inactive, slip systems (with Schmid factors $< 0.1$).
%
%
%
%
Therefore, in the three cases discussed so far, corresponding to the three regions near the corners of the stereographic triangle shown in Figure~\ref{fig:overall_stereo_behavior}, there is a clear separation between double- and single-exponential link length distributions on individual slip systems, and it is well correlated with the Schmid factors.
Slip systems with Schmid factors less than $0.1$ can be labeled as inactive slip systems and exhibit single-exponential link length distributions.

%

\subsection{\texorpdfstring{Loading axis near $[1\,2\,3]$}{Loading axis near [1 2 3]}}
\label{sec:near_123}

\begin{figure}[htbp]
\includegraphics[scale=1,center]{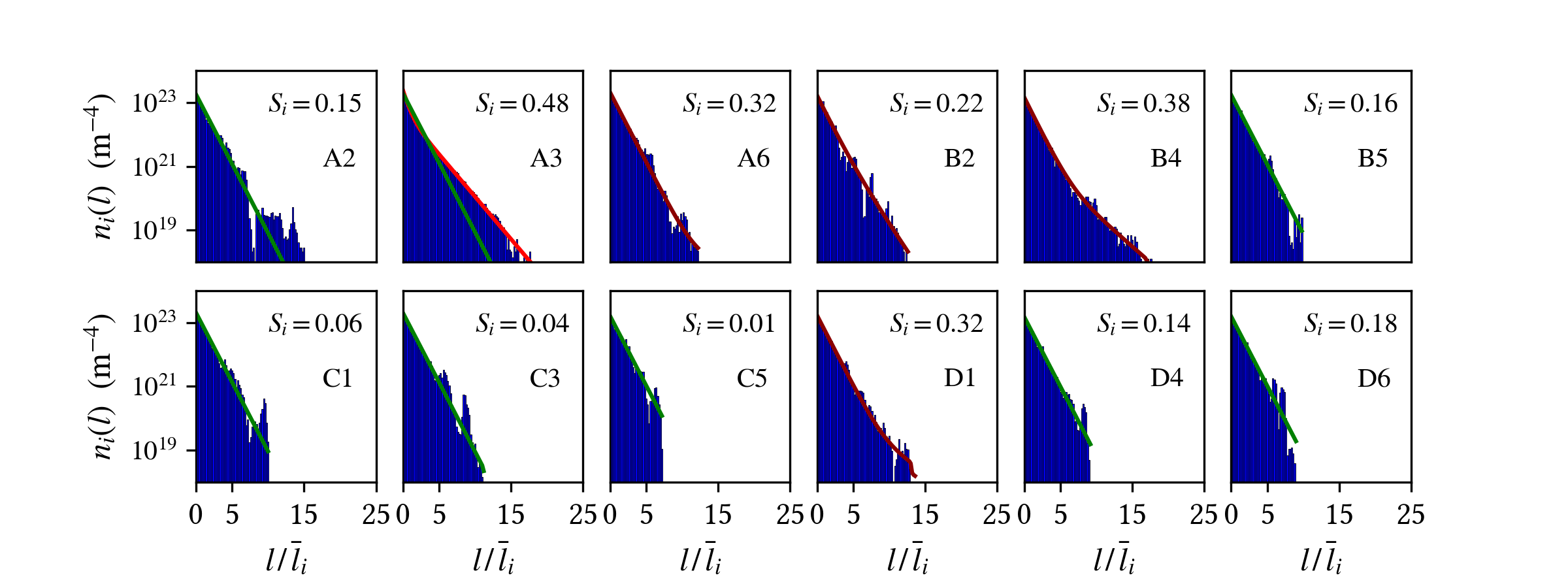}
\caption[justification=centering]{Link length distribution of dislocations on individual slip systems corresponding to a DDD simulation along $[0.24,\,0.49,\,0.84]$ loading orientation.}
\label{fig:LL_system_123}
\end{figure}



Figure~\ref{fig:LL_system_123} shows the link length distribution of $n_i(l)$ corresponding to loading orientation $[0.24, 0.49, 0.84]$
on the 12 slip systems. 
The observed behavior is typical of the loading orientations shown in the shaded area near the $[1\,2\,3]$ central region in Figure~\ref{fig:overall_stereo_behavior}.
It is interesting to note that the presence of double-exponential link distribution appears to be greatly reduced, compared with the three cases shown above.
Only the dominant slip system A3, with the highest Schmid factor (0.48), exhibits a clear double-exponential distribution.  Even here, the long tail in the double-exponential is far less prominent than those for active slip systems in the three cases shown above.
Because slip systems C1, C3, C5 have Schmid factors below 0.1, they are expected to be inactive and to have the single-exponential link distributions as observed.
However, slip systems A6, B4, D1 all have Schmid factors above 0.3; yet their link length distributions do not exhibit a clear double-exponential form.
This is in contrast to Figure~\ref{fig:LL_system_111} (for loading orientations near $[1\,1\,1]$) where slip systems with Schmid factors exceeding 0.2 all show a pronounced long tail in the double-exponential link length distribution.
Based on these observations, we hypothesize that having multiple slip systems with moderately high and comparable Schmid factors is important for the exhibition of a clear double-exponential link length distribution.  As a result, loading orientations near the three corners of the stereographic triangle is more conducive to double-exponential link length distribution than those near the center of the stereographic triangle.

%
%
%
%
%

\subsection{Statistical characteristics of long links}
\begin{figure}[htbp]
    \centering
    \includegraphics[width=0.45\linewidth]{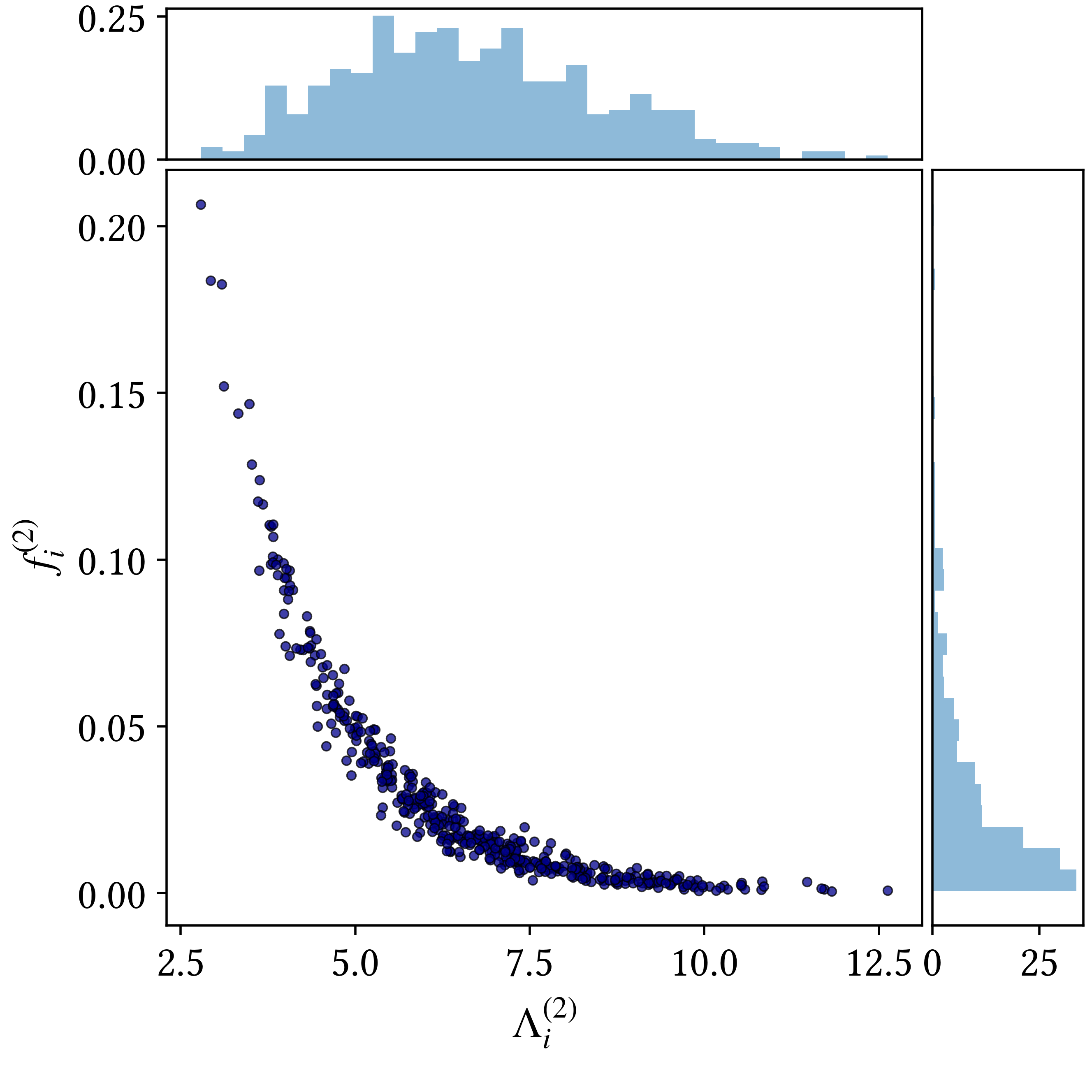}
    \caption{ The distribution of parameters $f^{(2)}_i$
    and $\Lambda^{(2)}_i$ 
    for all slip systems and different loading orientations when the link length distribution exhibits a clearly double-exponential distribution.
    $f^{(2)}_i$ measures the fraction of links belonging to the long tail and $\Lambda^{(2)}_i$ measures the average link length in the long tail normalized by the average link length on slip system $i$.
    }
    \label{fig:loading_orien_stats}
\end{figure}





We applied the link length analysis to dislocation configurations from DDD simulations along all 118 loading orientations.
For every loading orientation, the link-length distribution satisfies a double-exponential distribution on some slip systems, and a single-exponential on other slip systems.  Sometimes there are also slip systems on which the link length distribution appears somewhat in between these two behaviors, but cannot be well described by either one, such as those shown by a brown line in Figure~\ref{fig:LL_system_011} and Figure~\ref{fig:LL_system_123}.

For those slip systems on which a double-exponential form, Eq.~(\ref{eq:2Expo_i}), appears to be a good description of the link length distribution, we examine the statistical characteristics of the links in the long tail.
%
First, we introduce $l^\star$ as the transition length above which the link length distribution is dominated by the long tail.
Mathematically, $l^\star$ is defined as the length at which the two terms in Eq.~(\ref{eq:2Expo_i}) equal each other.  This leads to the expression
\begin{equation}
    l^\star = \left. \ln\left(\frac{N_i^{(1)}}{l_i^{(1)}}\cdot \frac{l_i^{(2)}}{N_i^{(2)}} \right) \, \right/ \, \left(\frac{1}{l_i^{(1)}} - \frac{1}{l_i^{(2)}}\right) .
\end{equation}
We then define $f^{(2)}_i$ as the fraction of the links belong to the long tail on slip system $i$,
\begin{equation}
  f^{(2)}_i
  \equiv \frac{1}{N_i} 
          \int_{l^\star}^{\infty}
          \frac{N_i^{(2)}}{\bar{l}_i^{(2)}}\,e^{-\,l/\bar{l}_i^{(2)}}\,dl \\
  = \frac{N_i^{(2)}}{N_i}  \,e^{-\,l^\star/\bar{l}_i^{(2)}}
\end{equation}
Finally, we define $\Lambda^{(2)}_i$ as the average length of the links in the long tail normalized by the average link length on slip system $i$.
\begin{equation}
  \Lambda^{(2)}_{i} \equiv \frac{1}{\bar{l}_i}
  \, \frac{1}{f^{(2)}_i \, N_i}
  \int_{l^\star}^{\infty}
         l \,\frac{N_i^{(2)}}{\bar{l}_i^{(2)}}\,e^{-\,l/\bar{l}_i^{(2)}}\,dl 
  = \frac{l^\star + \bar{l}_i^{(2)}}{\bar{l}_i} 
  = \frac{l^\star}{\bar{l}_i} + \lambda^{(2)}_i 
  = \lambda^{(2)}_i \left( 1 
    + \frac{l^\star}{\bar{l}_i^{(2)}}  \right) \, .
\end{equation}
%

%
Figure~\ref{fig:loading_orien_stats} shows the distribution of $f^{(2)}_i$ and $\Lambda^{(2)}_i$ parameters on all slip systems that clearly exhibited a double-exponential link length distribution.
It can be seen that in most cases, $f^{(2)}_i$ is between 0 and 0.05, meaning that the links belonging to the long tail typically constitute less than $5\%$ of all links.
In most cases, $\Lambda^{(2)}_i$ is between 5 and 10, meaning that the average link lengths in the long tail is about 5 to 10 times the average link length on the given slip system.  
Figure~\ref{fig:loading_orien_stats} also shows that $f^{(2)}_i$ and $\Lambda^{(2)}_i$ are negatively correlated, i.e. slip systems with large characteristic length $\Lambda^{(2)}_i$ tend to have smaller fractions of links in the long tail distribution.
Further research is needed to understand the physical meaning of such a correlation.

\section{Discussions}
\label{sec:discussion}

\subsection{Poisson process with growth for explaining link length distributions}\label{sec:2Expo_org}

Section~\ref{sec:near_001}
demonstrated (by relaxing the dislocation configurations to zero stress) that the applied stress during plastic deformation is responsible for producing the long tail in the link‑length distribution on active slip systems.
However, it did not explain why this distribution satisfies a double‑exponential form.
Here we provide an explanation through a simple model.
Previously, Sills et al.~\cite{sills2018dislocation} have shown that the (single) exponential distribution of link lengths can be explained by a one-dimensional Poisson process, in which a collection of links is randomly split (e.g., as a result of dislocation intersections).
We now show that incorporating link growth into this process can lead to double-exponential distributions.

Consider a collection of links whose lengths are $l_k$ ($l_k > 0$).
Every link $i$ can be randomly split into two links, $l_k^{(1)}$ (uniformly distributed from 0 to $l_k$) and $l_k^{(2)} = l_k - l_k^{(1)}$, at a probability rate $r_k$. 
This means that over an infinitesimal time period $dt$, a link with length $l_k$ will split with probability $r_k \, dt$.
In this model, we assume
\begin{equation}
    r_k = A\cdot \frac{l_k}{\bar{l}}
\label{eq:split_link}
\end{equation}
where $\bar{l}$ is the average link length (at the current time) in the collection, and $A$ is a constant (split rate coefficient).
Here we make the split rate $r_k$ proportional to the normalized link length ($l_k\,/\, \bar{l}$) so that $r_k$ can remain stationary even when $\bar{l}$ varies with time.

\begin{figure}[htbp]
    \centering
    \includegraphics[width=0.9\linewidth]{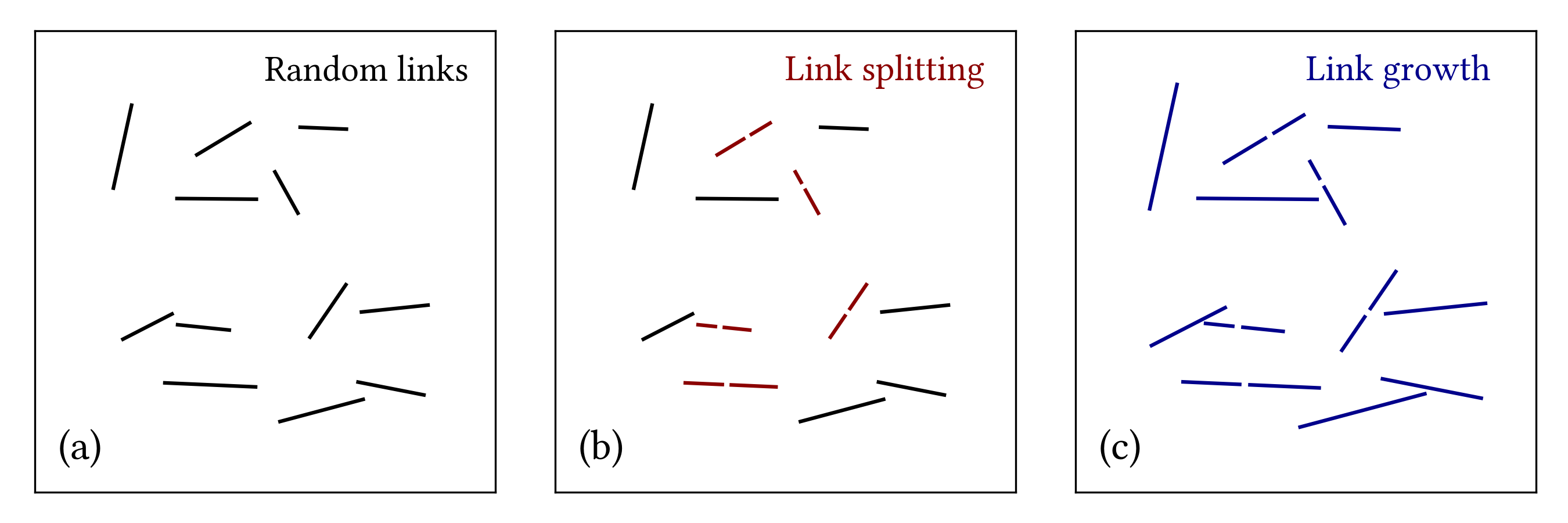}
    \caption{Schematic illustration of the generalized Poisson process.
(a) Initial distribution of links with random lengths.
(b) Some links are randomly split (shown in red), according to Eq.~\eqref{eq:split_link}.
(c) All links grow in length according to Eq.~\eqref{eq:growth_func}.}
    \label{fig:poisson_process_schematic}
\end{figure}

At the same time, every link also grows according to the following rule,
\begin{equation}
    {\dot{l}_k}\, / \,{l_k} = G \left({l_k}\, / \,{\bar{l}}\right)
\label{eq:growth_func}
\end{equation}
where $\dot l_k \equiv \frac{dl_k}{dt}$ is the growth rate of link $k$, and
$G(\cdot)$ specifies the relative growth rate as a function of the normalized link length.
The generalized Poisson process is illustrated in Figure~\ref{fig:poisson_process_schematic}.

\begin{figure}[htbp]
\centering
    (a)\includegraphics[width=0.35\linewidth]{Figures/l_ldot_constant.png}\hspace{20pt}(b)\includegraphics[width=0.35\linewidth]{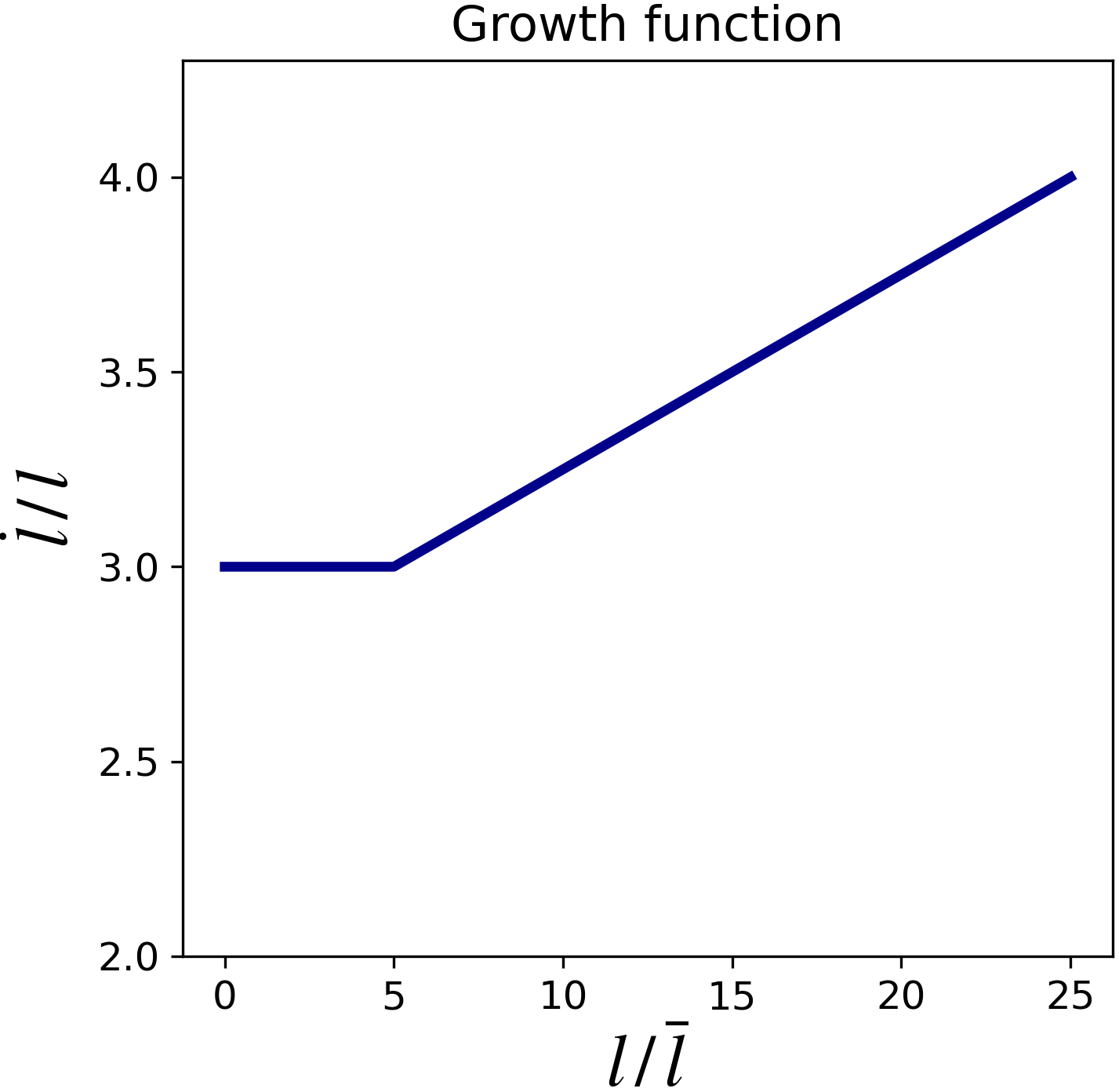}
    \\
    (c)\hspace{10pt}\includegraphics[width=0.32\linewidth]{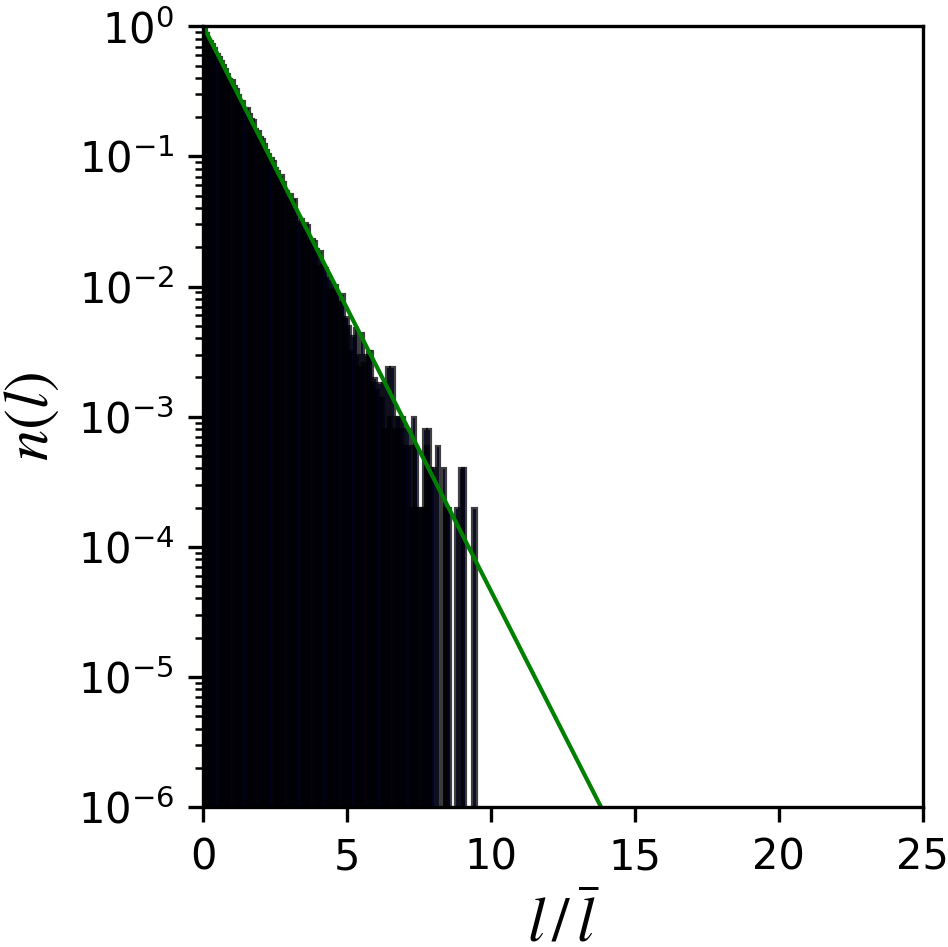}\hspace{20pt}
    (d)\hspace{10pt}\includegraphics[width=0.32\linewidth]{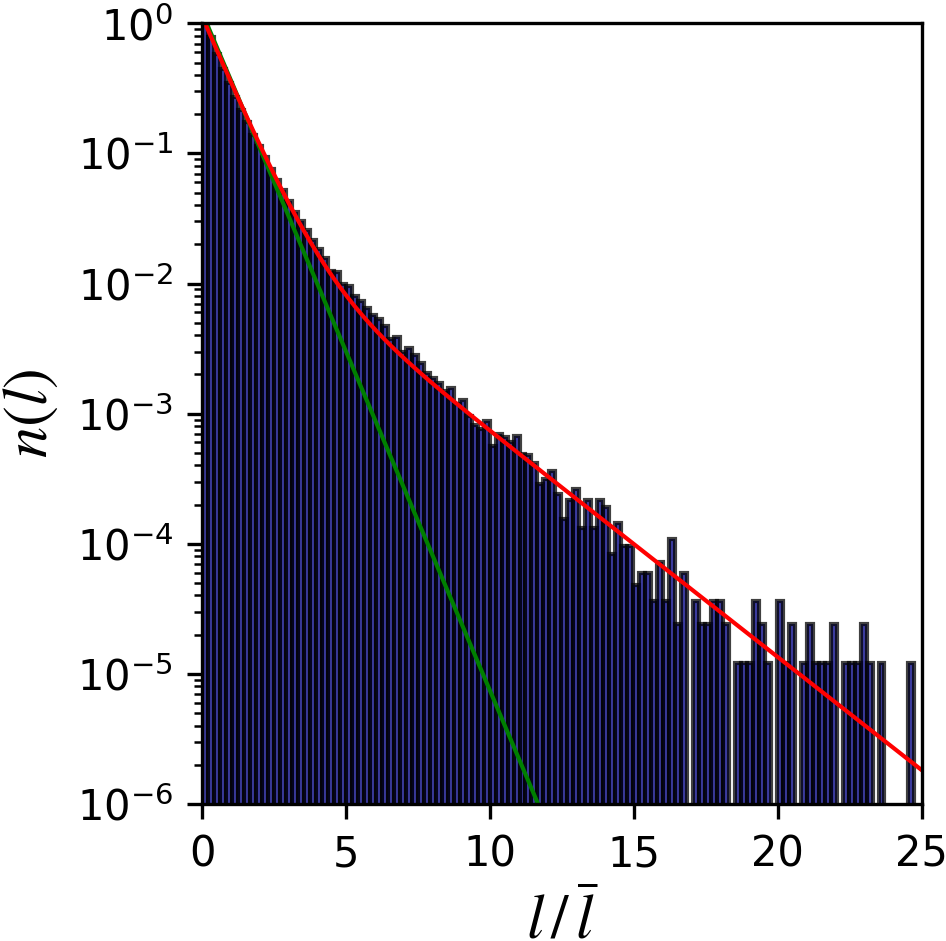}
    
    \caption[justification=centering]{
    Two model link growth functions 
    and their corresponding link length distribution. (a) A constant growth function $G = 3$ corresponding to a linear growth function ($\dot{l}$ proportional to $l$). (b) A piecewise linear growth function corresponding to super-linear growth beyond link threshold of $l\,/\,\bar{l} = 5$. (c) Link length distributions corresponding to the growth function in (a), exhibiting a single exponential. (d) Link length distributions corresponding to the growth function in (b), exhibiting a double-exponential distribution.}
    \label{fig:Li_dot}
\end{figure}

Figure~\ref{fig:Li_dot}(a) and (b) show two examples of the growth function $G$.  In Figure~\ref{fig:Li_dot}(a), $G$ is a constant which corresponds to a growth rate $\dot{l}$ that is linear in link length. 
In Figure~\ref{fig:Li_dot}(b), $G$ is a constant for $l\,/\,\bar{l}$ less than a threshold and a linear function of $l\,/\,\bar{l}$ beyond the threshold, leading to a growth rate $\dot{l}$ that contains a quadratic term as a function of link length (super-linear growth).
Note that in a real crystal the dislocation link growth rate depends on many factors due to the interaction with other dislocations, and dislocation links of the same length do not necessarily grow at the same rate. However, in order to provide a simple explanation for the origin of the double exponential distribution, here we take a model in which the link growth rate is only a function of link length.

We numerically implemented the extended Poisson process model with $A = 1$ in Eq.~\eqref{eq:split_link}.
The simulation begins with 1000 links sampled from a uniform distribution, $l \sim U(0,\, l_\text{max})$, where $l_\text{max} = 10$ (in arbitrary units).
The evolution of the link collection is simulated using a timestep of \revisions{$\Delta t = 5\times10^{-5}$}.
At each timestep, we compute the split probability using $r_k\Delta t$ for every link $i$, and split it into two links with this probability.
As a self-consistency check, we have verified that $r_k \Delta t < 0.01$ for all links throughout all the simulations.
Every link also grows in length according to the growth function at every time step.
We note that the number of links in the simulation does not remain constant due to the splitting events.
To keep the computational cost and memory requirement under control, we impose an upper limit on the number of links in our collection.
When the number of links exceeds \revisions{$N_{\rm max} = 10^6$}, we resample the collection by randomly choosing $N_{\rm max}\, /\, 3$ links to keep and removing the rest.
It is observed that as the simulation proceeds, the normalized link length distribution evolves toward a steady state for both the growth functions shown in Figure~\ref{fig:Li_dot}.
%

Figure~\ref {fig:Li_dot}(c) shows the limiting distribution of link lengths normalized by the average length corresponding to the growth function shown in Figure~\ref{fig:Li_dot}(a).
In this case, $G$ is a constant, which means that the links grow at a rate $\dot{l}$ proportional to their current length $l$.
The resulting link length distribution follows a single exponential function defined in Eq.~\eqref{eq:1Expo_i}.
On the other hand, Figure~\ref{fig:Li_dot}(d) shows the limiting distribution of normalized link length corresponding to the growth function shown in Figure~\ref{fig:Li_dot}(b).
Here $G$ is a constant (3) for $l\,/\,\bar{l}\le 5$ but becomes a linear function (greater than 3) of $l\,/\,\bar{l}$ when the latter exceeds $5$.
The resulting link length distribution follows a double exponential function defined in Eq.~\eqref{eq:2Expo_i}.
With the parameters chosen in this model, the double-exponential distribution can be characterized by the parameters $\lambda_i^{(1)} = 0.83$ and $\lambda_i^{(2)} = 2.50$.
These values are close to the fitted parameters for link length distributions on active slip systems in Table~\ref{tab:2exp_ddd_fit}. 
\revisions{We note that it is not necessary for the growth function to exhibit a sharp corner, as shown in Fig.~\ref{fig:Li_dot}(b), in order for the steady-state link length distribution to exhibit the double-exponential behavior.
%
%
%
As long as the transition of the slope of the growth function occurs over a narrow range of $x=l/\bar{l}$, qualitatively the same steady-state link length distribution will be produced.
For example, if we use $G(x)=\frac{k}{\alpha}\ln\left[1+{\rm e}^{\alpha(x-x_0)}\right]+\beta$, with $k = 0.05$, $\alpha = 1$, $x_0 = 5$, and $\beta = 3.00$, then the resulting steady-state link length distribution will look essentially the same as Fig.~\ref{fig:Li_dot}(d).
%
%
%
This simple model also predicts that the steady-state distribution is well-reached by the time each link has split by about 10 times on average.
}

In summary, we have shown that both the single- and double-exponential distributions of link lengths can result from an extended Poisson process depending on the specific form of the growth function $G$.
We note that the link length distribution always goes to a single-exponential as long as the relative growth rate $G$ is a constant, regardless of the value of this constant (including zero).
For the long tail in the double-exponential to appear, $G$ has to be greater than a constant for link lengths exceeding a threshold.
Physically, this super-linear growth corresponds to the bowing out of long links on active slip systems driven by the applied shear stress.
Our results support the picture that long links on active slip systems behave differently from shorter links in that the long links grow much faster.

\subsection{Correlation between link length and link velocity}

\label{sec:link_velocity}
\begin{figure}\centering
\includegraphics[width=0.45\linewidth]{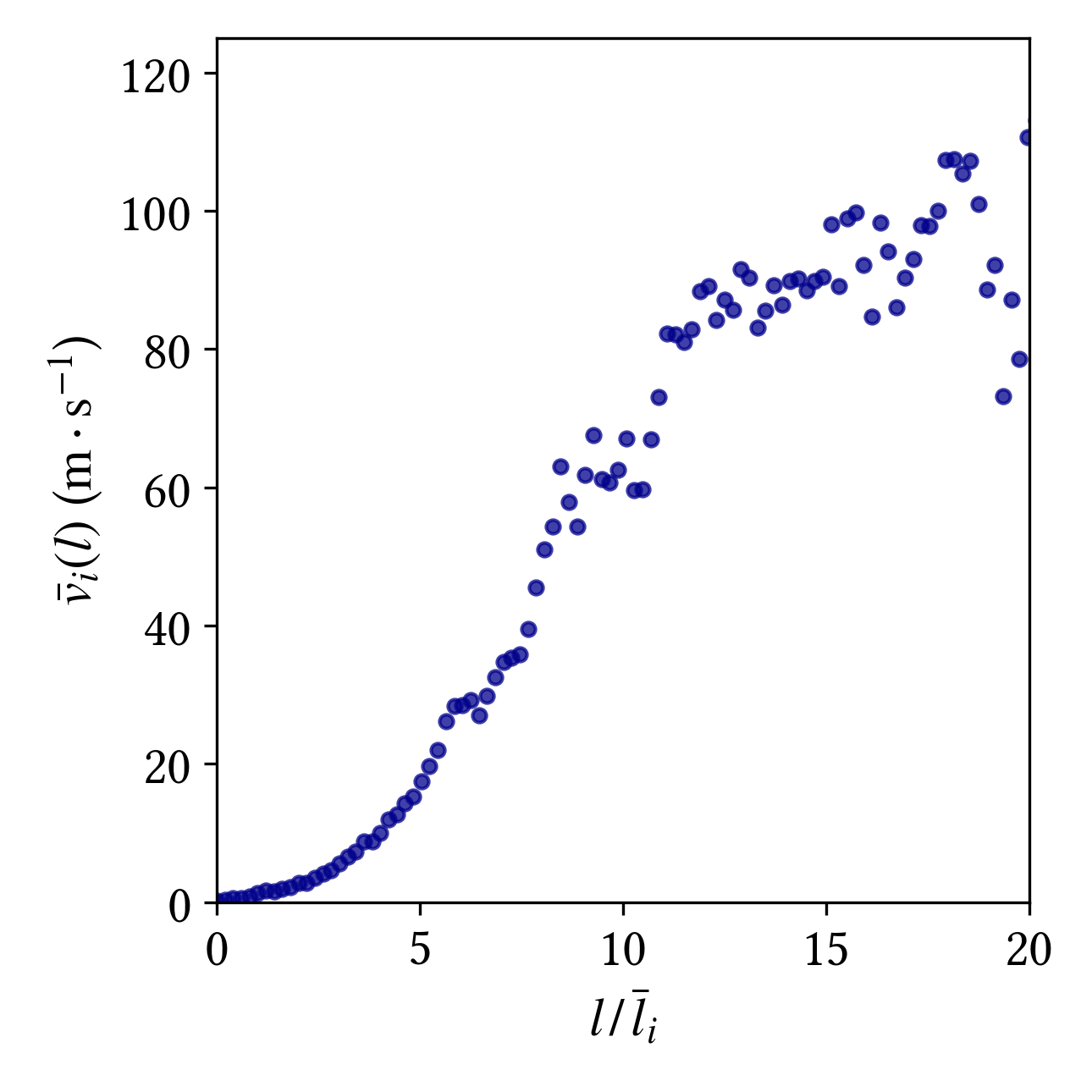}
    \caption{
    Average link velocity as a function of regards to link length on slip system A3 for loading direction $[0.03,\,0.05,\,0.99]$.  The averaging is over the same strain range as shown in Figure~\ref{loading schematic}.}
    \label{fig:vel_dist_linklen}
\end{figure}

We hypothesized in Section~\ref{sec:near_001}
that long links exhibit higher velocities, contributing to the long-tail behavior of the second exponential distribution.
Figure~\ref{fig:vel_dist_linklen} shows the averaged link velocities on slip system A3 as a function of link length, providing evidence supporting this hypothesis.
The average link velocity is defined as $\bar{v}_i(l) = \dot{\gamma}_i (l)\,/\, \left(l \, b \, n_i(l)\right)$, where $\dot{\gamma}_i(l)dl$ is the 
contribution to the shear strain rate on slip system $i$ for links with length within $[l, l+dl)$, 
and $b$ is the magnitude of the Burgers vector.
The loading orientation is $[0.03, 0.05, 0.99]$ and the data is extracted over the same set of dislocation configurations that lead to the link length distribution in Figure~\ref{fig:LL_system}.
We observe that the average link velocity remains relatively low (less than $20\, {\rm m}\cdot{\rm s}^{-1}$)
when $l\,/\,\bar{l}_i < 5$.
Links with $l\,/\,\bar{l}_i$ exceeding 5 exhibit a pronounced increase in average velocity with length, with velocity reaching $100\, {\rm m}\cdot{\rm s}^{-1}$ for $l\,/\,\bar{l}_i > 15$.
%
%
%
%




\section{Conclusions\label{sec:conclusion}}



Our analysis of Discrete Dislocation Dynamics (DDD) simulation data across various loading orientations reveals that link length distributions on individual slip systems under applied load may exhibit either a single- or a double-exponential form. Specifically, active slip systems display a distribution characterized by a sum of two exponential components, while inactive slip systems conform to a single-exponential distribution. We hypothesize that the distinctive long tail in the double-exponential distribution on active slip systems originates from the bowing out of longer links under applied stress. This hypothesis is supported by the observation that, upon relaxation of the applied stress, the distribution reverts to a single-exponential form.
The DDD data also show that during the plastic deformation, the average velocity of longer links is greater than that of shorter links.

To explain the origin of the observed distributions, we propose a generalized Poisson process model with both splitting and growth, and show that it can reproduce both the single- and double-exponential link length distributions. A linear growth rate (i.e., constant $\dot{l}\,/\,l$) results in a single-exponential distribution. In contrast, a super-linear growth rate for links exceeding a critical length leads to a double-exponential distribution. This indicates that the accelerated growth of longer links provides the underlying mechanism for the long tail, reinforcing our hypothesis that stress-induced link bowing is responsible for the emergence of this double-exponential behavior.

Analysis across all 118 loading orientations shows that link length distributions near the corners of the stereographic triangle (i.e., highly symmetric loading orientations) exhibit clear distinctions between single- and double-exponential link length distributions.
%
In these cases, this distinction is clearly correlated with the Schmid factor.
%
%
In contrast, in loading directions near the center of the stereographic triangle, the long tail in the double-exponential distribution appears greatly suppressed, and the distinction between single- and double-exponential distributions are sometimes not very clear.
%
For slip systems that do exhibit a clear double-exponential link length distribution, the parameters $f^{(2)}$ and $\lambda^{(2)}$ characterizing the long tail exhibit a negative correlation with each other.
%
This suggests that longer characteristic lengths in the long-tail often lead to fewer links in the long-tail population.
%

Our results demonstrate the promise of extracting physical insights from analyzing the microstructural data from DDD simulations.
This work contributes to the broader effort of systematically coarse-graining data from DDD simulations to develop physics-based constitutive relations. While our current focus is solely on the geometric characteristics of the dislocation network, future work will aim to establish quantitative connections between link length, network geometry, and the resulting slip rates, thereby advancing towards a comprehensive theory for predicting slip behavior of crystals based on dislocation dynamics.

\section*{Declaration of Competing Interest}

None.

\section*{CRediT author statement}


{\bf Sh.\ Akhondzadeh}: Conceptualization, Methodology, Software, Validation, Data curation, Writing - original draft, Writing -- review \& editing.
{\bf Hanfeng Zhai}: Conceptualization, Methodology, Software, Validation, Visualization, Data curation, Writing -- original draft, Writing -- review \& editing.
{\bf Wurong Jian}: Validation, Writing -- review \& editing.
{\bf Ryan B. Sills}: Conceptualization, Writing -- review \& editing.
{\bf Nicolas Bertin}: Conceptualization, Writing -- review \& editing.
{\bf Wei Cai}: Conceptualization, Methodology, Software, Validation, Writing -- review \& editing. Supervision.

\section*{Acknowledgment}


This work was supported by the U.S. Department of Energy, Office of Basic Energy Sciences, Division of Materials Sciences and Engineering under Award No. DE-SC0010412 (Sh. A. and W. C.).
Research was sponsored by the Army Research Office, United States, and was
accomplished under Grant Number W911NF-21-1-0086 (R. B. S.).
N.B.'s work was performed under the auspices of the U.S. Department of Energy by Lawrence Livermore National Laboratory under Contract DE-AC52-07NA27344.

\section*{Data availability}

Related code and data are to be released on GitHub upon acceptance of the manuscript.

\appendix

\renewcommand{\thefigure}{A\arabic{figure}} 
\setcounter{figure}{0}  
\renewcommand{\theequation}{A\arabic{equation}} 
\setcounter{equation}{0}  
\renewcommand{\thetable}{A\arabic{table}} 
\setcounter{table}{0}  






\section*{Appendix}

\section{DDD simulation details}
\label{app:ddd_details}

%
The material properties of copper single crystal used in the DDD simulations are shown in Table~\ref{tbl:simParam}. 
Simulations were performed using an initial configuration with the dislocation density of $\rho_0\approx1.2\times 10^{12} \rm{m}^{-2}$. 
This value is twice as high as that used in Ref.~\cite{sills2018dislocation}, resulting in more links in the simulation volume.

\begin{table}[htbp]
\centering
\small
\caption{Summary of simulation parameters. For more details, see Ref.~\cite{JMPS1}. } \label{tbl:simParam}
\begin{tabular}{ P{5cm}  P{1.6cm}  P{2.9cm}  }
\toprule
Property & Parameter & Value  \\
\toprule
Shear modulus & $\mu$ & 54.6 GPa  \\
Poission's ratio & $\nu$ & 0.324   \\
Burgers vector magnitude & $b$ & 0.255 nm   \\
Drag coefficient & $B$ & 15.6 $\mu\rm Pa\cdot s $  \\
 \midrule
 \bottomrule
\end{tabular}
\end{table}

\clearpage
\printbibliography

\end{document}